# Quantum Mechanics as Complex Probability Theory

Saul Youssef

Supercomputer Computations Research Institute
Florida State University
Tallahassee, Florida 32306–4052
youssef@scri.fsu.edu

**Abstract**

Realistic quantum mechanics based on complex probability theory is shown to have a frequency interpretation, to coexist with Bell's theorem, to be linear, to include wavefunctions which are expansions in eigenfunctions of Hermitian operators and to describe both pure and mixed systems. Illustrative examples are given. The quantum version of Bayesian inference is discussed.

# 1. Introduction

Quantum mechanics can be reformulated as a realistic theory where quantum effects arise from a breakdown of conventional probability theory rather than from the usual wave–particle duality. Consider, for example, the two slit experiment where a source emits a single particle which travels towards a wall with two slits and is detected at position $x$ on a screen located behind the wall. The usual argument concluding that an interference pattern on the screen implies that the particle did not either go through one slit or the other is ultimately an argument in probability theory where

$$P(x) = P(x \text{ via slit } 1) + P(x \text{ via slit } 2) \tag{1}$$

is the critical statement. As an alternative, it is possible to consider a realistic theory, where the particle does go through one slit or the other, but where (1) fails due to a failure of probability theory itself. The basic consequences of this idea and its consistency with Bell's theorem and with other limitations on local realism are discussed in references 1 and 2. Here we develop this approach in more detail with emphasis on insights which are not available in standard quantum mechanics.

# 2. Probability Theory as Physics

Probability theory is often introduced as a theory of experiments which may succeed or may fail due to some random influence. If an experiment is successful $n$ times in $N$ trials, one calls the large $N$ limit of $n/N$ the "probability of success." Probability theory then results from assuming that these probabilities follow Kolmogorov's axioms.[3] For our purposes, however, it is convenient to adopt the more general Bayesian view where, given a pair of propositions $(a, b)$, probability $P(a, b)$ gives a non–negative real number meant to measure how likely it is that $b$ is true if $a$ is known.[4] One can then follow Cox[5] and assume that $P(a, b \wedge c) = F(P(a, b), P(a \wedge b, c))$ and $P(a, \neg b) = G(P(a, b))$ for some fixed functions $F$ and $G$ and derive the Bayesian form of probability theory. A function $P$ mapping pairs of propositions into the interval [0,1] is a *probability* if

(a) $P(a, b \wedge c) = P(a, b)P(a \wedge b, c)$, (2)
(b) $P(a, b) + P(a, \neg b) = 1$, (3)
(c) $P(a, \neg a) = 0$ (4)

for all propositions $a$, $b$ and $c$. In its Bayesian form, probability theory is just a consistent way of assigning a "likelihood" in the interval [0,1] to any pair of propositions whether or not these propositions refer to the outcome of a random experiment.[6] As we will see, the Bayesian view is essential for constructing and understanding the predictions of our extended probability theory.

To derive a frequency interpretation for ordinary probabilities, let $p$ be the probability of success in an experiment and note that by the central limit theorem, the number of successes $n$ in $N$ independent copies of the experiment is asymptotically gaussian with mean $\mu = Np$ and with $\sigma/\mu$ proportional to $1/\sqrt{N}$. The probability for $n/N$ to be in any interval not containing $p$ can then be made arbitrarily small by increasing $N$. Thus, we have a frequency interpretation provided that we assume that an arbitrarily small probability for $n/N$ to be in some interval means that $n/N$ is never actually observed to be in that interval, or, more generally, that $P(a, b) = 0$ means that $b$ is never observed if $a$ is known to be true. As innocent as it seems, this additional property brings probability theory into the domain of physics and leads to a question: does such a $P$ exist? If we assume that there is no conventional probability $P$ with this additional property, Cox's remaining axioms may still be preserved provided probabilities are allowed to be complex.[7] Since Bayesian probabilities are not defined as frequencies, it is possible to consider complex probabilities provided a consistent frequency interpretation can be constructed after the fact (section 3). In the complex case, Cox's arguments follow as before and the resulting complex probability theory has exactly the same form as (2)–(4), only with a complex $P$. In order to distinguish complex probabilities from the conventional case, an arrow notation is used so that the complex probability that $b$ is true given that $a$ is known is denoted by "$(a \to b)$" rather than by "$P(a, b)$." The arrow is meant to suggest a numerical version of implication.

We are now ready to attempt a description of quantum phenomena by combining complex probability with a simple definition of realism. An arrow function mapping pairs of propositions into the complex numbers is a *quantum theory* if

I. The arrow is a *complex probability*:

$$(a \to b \wedge c) = (a \to b)(a \wedge b \to c) \tag{I.a}$$
$$(a \to b) + (a \to \neg b) = 1 \tag{I.b}$$
$$(a \to \neg a) = 0 \tag{I.c}$$

for all propositions $a$, $b$ and $c$ and if

II. There exists a set $U$ (the *state space*) and a real *time* $t$ such that:

$$x_t \wedge y_t = false \text{ if } x \neq y \tag{II.a}$$
$$(a_t \to b_{t''}) = (a_t \to U_{t'} \wedge b_{t''}) \tag{II.b}$$
$$(a_t \wedge x_{t'} \to b_{t''}) = (x_{t'} \to b_{t''}) \tag{II.c}$$

for all $x, y \in U$, for all propositions $a, b$ and for all times $t \leq t' \leq t''$ with $U_t \equiv \vee_{x \in U} x_t$.

The realistic state axioms (II.a–II.c) guarantee that a system cannot be in two states at once (II.a), that a system is in some state at each intermediate time (II.b) and that the knowledge that a system is in some particular state makes all previous knowledge irrelevant (II.c). For the purposes of this paper, it is convenient to add the following technical assumptions: III. $U$ is a measure space and for any times $t \leq t'$ for any proposition $a_t$, $(a_t \to U_{t'}) = \int_{x \in U}(a_t \to x_{t'})$ and $(a_t \to U_{t'})$ is differentiable with respect to $t'$. Note that unlike the "state space" of conventional quantum mechanics, $U$ is not a Hilbert space.

Many simple facts from probability theory also follow in quantum theories. For all propositions $a$, $b$ and $c$, $(a \to a) = 1$, $(a \to true) = 1$, $(a \to false) = 0$ and $(a \to b \vee c) = (a \to b) + (a \to c) - (a \to b \wedge c)$ and, if $(a \to b) \neq 0$ then $(a \wedge b \to c) = (a \wedge b \to b \wedge c)$. It is convenient to call a proposition $e_t$ *normal* if $\int_{x \in U} |e_t \to x_{t'}|^2$ exists and is greater than zero for all $t' \geq t$ and to let a time subscript on a set of propositions denote the *or* of all its elements: $W_t = \vee_{w \in W} w_t$. Following probability theory, propositions $a$ and $b$ are said to be *independent* if $(q \wedge a \to b) = (q \to b)$ for all propositions $q$. The rest of this work is an exploration of the consequences of I-III.

## 3. Reconstructing a Frequency Interpretation

As discussed above, the physical meaning of conventional probability is fixed by the assumption that if proposition $a$ is known and if $P(a,b) = 0$, then $b$ never happens in real experiments. Since, however, complex probabilities may sum to zero without being zero individually, we make this assumption for $(a \to b)$ only if $b = x_t$ for some $x$ in the state space. More generally, if $U$ is a quantum theory, $t \leq t'$, $e_t$ is a normal proposition and $a$ and $b$ are any propositions, define a function *Prob* by

$$Prob(e_t, a_{t'}) = \frac{\int_{x \in U} |e_t \to a_{t'} \wedge x_{t'}|^2}{\int_{x \in U} |e_t \to x_{t'}|^2} \qquad (5)$$

and

$$Prob(e_t, a_{t'} \wedge b_{t'}) = \frac{\int_{x \in U} |e_t \to a_{t'} \wedge b_{t'} \wedge x_{t'}|^2}{\int_{x \in U} |e_t \to x_{t'}|^2} \qquad (6)$$

where, under conditions described below, *Prob* will be shown to have many of the properties of a conventional probability and, in particular, provides a frequency interpretation for quantum theories.

Supposing that $U$ is a quantum theory and $W$ is a set of propositions, we say that $W$ *supports probabilities with initial knowledge* $e_t$ if $e_t$ is normal and if, for all $t' \geq t$, for all $A, B \subset W$,

(a) $A_{t'} \wedge A^c_{t'} = false$ and $U_{t'}$ implies $W_{t'}$ \hfill (7)
(b) $Prob(e_t, A_{t'} \wedge B_{t'}) = Prob(e_t, A_{t'})\, Prob(e_t \wedge A_{t'}, B_{t'})$ \hfill (8)
(c) $Prob(e_t, A_{t'}) + Prob(e_t, A^c_{t'}) = 1$ \hfill (9)

where the complement $A^c = W - A$. Following the standard argument sketched in the previous section, if $Prob$ has these properties for a set of propositions $W$, then the probabilities $Prob(e_t, A_{t'})$ have a frequency interpretation. The conditions under which $W$ supports probabilities can be easily found.

Theorem 1 (Orthogonality). *If $U$ is a quantum theory, $e_t$ is a normal proposition and $W$ is a set of propositions satisfying point a) above, then $W$ supports probabilities with initial knowledge $e_t$ if and only if, for all times $t' \geq t$ and for all subsets $A$ of $W$, $\int_{x \in U}(e_t \to A_{t'} \wedge x_{t'})^*(e_t \to A_{t'}^c \wedge x_{t'}) + (e_t \to A_{t'} \wedge x_{t'})(e_t \to A_{t'}^c \wedge x_{t'})^* = 0$.*

*Proof.* We need to show that $W$ supports probabilities with initial knowledge $e_t$. Condition (a) above is already assumed, (b) follows easily from the definition of $Prob$, and (c) follows since $(e_t \to x_{t'}) = (e_t \to W_{t'} \wedge x_{t'}) = (e_t \to A_{t'} \wedge x_{t'}) + (e_t \to A_{t'}^c \wedge x_{t'})$ and therefore $1 = \int_U |e_t \to x_{t'}|^2 / \int_U |e_t \to x_{t'}|^2 = Prob(e_t, A_{t'}) + Prob(e_t, A_{t'}^c) + I(A) / \int_U |e_t \to x_{t'}|^2$ where $I(A)$ is the integral in the theorem. Similarly for the reverse implication.

Since the state space $U$ of a quantum theory trivially satisfies the conditions of theorem 1, $U$ supports probabilities with any normal initial knowledge, and given any $A \subset U$, $Prob(e_t, A_{t'})$ has a proper frequency interpretation and thus predicts how often a system will be found in region $A$ at time $t'$ given that $e$ is known at time $t$.

It is useful to consider extending the definition of $Prob$ to mixed times. If, for normal $e_t$ with $t \leq t' \leq t''$ we were to define

$$Prob(e_t, A_{t'} \wedge B_{t''}) = \frac{\int_{x \in U} |e_t \to A_{t'} \wedge B_{t''} \wedge x_{t''}|^2}{\int_{x \in U} |e_t \to x_{t''}|^2} \tag{10}$$

then

$$Prob(e_t, A_{t'} \wedge B_{t''}) = \frac{\int_U |e_t \to A_{t'} \wedge x_{t''}|^2}{\int_U |e_t \to x_{t''}|^2} Prob(e_t \wedge A_{t'}, B_{t''}) \tag{11}$$

which does not generally satisfy (2) unless $t' = t''$. In particular, $\int_{x \in U} |e_t \to A_{t'} \wedge x_{t''}|^2 = \int_{x \in U} |\int_{y \in U}(e_t \to A_{t'} \wedge y_{t'})(y_{t'} \to x_{t''})|^2$ and, if one could take the square inside the integral (and assuming $Prob(y_{t'}, U_{t''}) = 1$), (2) would be satisfied. Since this fails because of the "interference terms" involved in exchanging the square with the integral, this suggests that an appropriate classical limit would restore probability theory (2)–(4) for propositions with mixed times. In addition, this failure demonstrates the reason that Bell's theorem does not rule out quantum theories in spite of the fact that they are realistic and local. In Bell's analysis,[8] two spin $\frac{1}{2}$ particles in a singlet state are emitted towards two distant Stern–Gerlach magnets. Let $e_t$ define the known orientations of the two magnets and the description of the initial singlet state and let $M_{t''}$ be a description of one of the possible results of the final measurements. Let $t$ be the time when the singlet state is released, $t''$

be the time of the final measurement and let $t < t' < t''$. Bell's argument is an argument in probability theory beginning with an expansion in "hidden variable" $\lambda$ in state space $U$: $P(e_t, M_{t''}) = P(e_t, U_{t'} \wedge M_{t''})$ and so

$$P(e_t, M_{t''}) = \int_{\lambda \in U} P(e_t, \lambda_{t'} \wedge M_{t''}) = \int_{\lambda \in U} P(e_t, \lambda_{t'}) P(e_t \wedge \lambda_{t'}, M_{t''}) \qquad (12)$$

which, from our point of view, fails in the last step since $t' \neq t''$. Thus, although Bell's theorem is usually interpreted as ruling out local realistic theories, in a more general context Bell's result actually shows that one must choose between local realism and conventional probability theory. Discussion of other limitations on local realistic theories can be found in reference 2.

## 4. Examples of Quantum Theories

In order to provide some examples, it is helpful to begin the process of classifying quantum theories by their global properties. For example, one expects many quantum theories to be *time invariant* in the sense that for any propositions $a$ and $b$, $(a_t \to b_{t'}) = (a_{t+\tau} \to b_{t'+\tau})$ for all times $t, t', \tau$. If such a quantum theory is also *conservative* in the sense that $U_{t'}$ implies $U_t$ for all $t \leq t'$, then we have

Theorem 2. *If a quantum theory $U$ is conservative and time invariant, then there exists a function $\phi : U \to C$ and a complex number $\lambda$ such that $(U_t \to x_{t'}) = e^{\lambda(t'-t)} \phi(x)$ for all $x \in U$ and for all times $t < t'$.*

*Proof.* Let $t < t' < t''$ and $f(\tau) = (U_t \to U_{t+\tau})$. Then using the axioms and the conservative property, $(U_t \to U_{t''}) = (U_t \to U_{t'} \wedge U_{t''}) = (U_t \to U_{t'})(U_{t'} \to U_{t''})$ and so $f(a+b) = f(a)f(b)$ for all real $a, b > 0$. Since $f$ is differentiable (by III), $f(a) = e^{\lambda a}$ for some complex $\lambda$ and $(U_t \to x_{t'}) = (U_t \to U_{t'} \wedge x_{t'}) = (U_t \to U_{t'})(U_{t'} \to x_{t'})$ and the theorem is proved if we let $\phi(x) = (U_{t'} \to x_{t'})$ which is time independent by the time invariance property.

In the conventional view of quantum mechanics, theorem 2 is puzzling. After assuming very little about the system, one has concluded that the wavefunction is an energy eigenstate $\Psi(x, t) = e^{\lambda t} \phi(x)$. What if the system is, in fact, not in an energy eigenstate but is instead in a superposition of two different energy eigenstates? To answer this question, recall that a complex probability $(a_o \to b_t)$ is the best estimate of the truth of $b_t$ given that $a_o$ is known. However, if some additional fact $X$ is also known, one should instead calculate $(a_o \wedge X \to b_t)$. Thus, a wavefunction such as $\Psi$ should not be mistaken for "the state of the system." $\Psi$ merely represents the best estimate of the complex probabilities for $x_t (x \in U)$ given that only $U_o$ is known at $t = 0$. Since different observers may have different knowledge about a system, they may also describe a single system with different wavefunctions. This also implies that if an observer does not know all the relevant facts about a system, the corresponding wavefunction may give incorrect predictions. This, however, is not a failure of quantum theory any more than it is a failure of probability theory when the usual analysis of throwing a die fails in the case of a loaded die. In both cases, the theories are successful only if the relevant facts are known. To answer the original question, it is not possible for a superposition of two energy eigenstates to be the optimal description of the system unless some additional fact $X$ is known in addition to $U_o$. To take another example, consider a particle confined in a box containing some known but time invariant potential and where one initially knows only that the particle is somewhere in the left half of the box. One expects that, typically, this initial information will be become less useful as time goes on until finally, the initial knowledge has no influence on the optimal wavefunction. This suggests that we call a quantum theory *forgetful* if, for all propositions $a, b, c$ and for all times $t$, $(a_t \wedge b_{t'} \to c_{t''}) = (b_{t'} \to c_{t''})$ in the limit $t', t'' \to \infty$. As might be expected, you can easily show that forgetful, conservative, time invariant quantum theories evolve into energy eigenstates independent of the initial knowledge of the system.

For an example with more detailed predictions, consider a conservative, time invariant quantum theory $U$ with a *trap* $\beta \subset U$ in the sense that $\beta_t \Rightarrow \beta_{t'}$ for all $t \leq t'$. Since $U$ is conservative, $\alpha_{t'} \Rightarrow \alpha_t$ where $\alpha \equiv U - \beta$ [since $\alpha_{t'} \Rightarrow U_{t'} \Rightarrow U_t = (\alpha_t \vee \beta_t)$]. By arguments similar to the proof of theorem 2, one can show that for some complex $\lambda, k_o$, $(\alpha_o \to \alpha_t) = e^{\lambda t}$ and $(\alpha_o \to \beta_t) = k_o(1 - e^{\lambda t})$. Given that the system is not initially trapped, the probability to remain outside the trap for an additional time $t$ is

$$Prob(\alpha_o, \alpha_t) = \frac{\int_{x \in U} |\alpha_o \to \alpha_t \wedge x_t|^2}{\int_{x \in U} |\alpha_o \to x_t|^2} = \frac{\int_{x \in \alpha} |\alpha_o \to x_t|^2}{\int_{x \in \alpha} |\alpha_o \to x_t|^2 + \int_{x \in \beta} |\alpha_o \to x_t|^2} \quad (13)$$

and since $\int_\alpha |\alpha_o \to x_t|^2 = \int_\alpha |\alpha_o \to \alpha_t \wedge x_t|^2 = |e^{\lambda t}|^2 \int_\alpha |\alpha_t \to x_t|^2$ and similarly $\int_\beta |\alpha_o \to x_t|^2 = \int_\beta |\alpha_o \to \beta_t \wedge x_t|^2 = |k_o(1 - e^{\lambda t})|^2 \int_\beta |\alpha_o \wedge \beta_t \to x_t|^2$ and so

$$Prob(\alpha_o, \alpha_t) = \frac{1}{1 + k(t)|e^{-\lambda t} - 1|^2} \quad (14)$$

where $k(t) = |k_o|^2 \int_\beta |\alpha_o \wedge \beta_t \to x_t|^2 / \int_\alpha |\alpha_t \to x_t|^2$. If, in addition, $U$ is forgetful, then $k(t)$ is time independent in the large $t$ limit and, assuming that $\lambda$ is real and negative, results in the usual exponential decay law. If $\beta$ is a single point in the state space then, using II.c, $k(t)$ is time independent for all $t$ and if $t$ is real and negative, has the expected large $t$ behavior.[9]

As an even more detailed example, one can extract the complete dynamics of a scalar particle in a time invariant quantum theory with a state space $U = R^d$. Since, for any initial proposition $e_o$, the wavefunction $(e_o \to x_t)$ is given by $(e_o \to x_t) = (e_o \to U_o \wedge x_t) = \int_{y \in U} (e_o \to y_o)(y_o \to x_t)$, the time development of any initial wavefunction is determined by the "propagator" $(x_t \to x'_{t'})$. One can now construct a path integral by choosing times $t_o < t_1 < t_2 < \ldots < t_n$ with $t_o = t$ and $t_n = t'$. Letting $x_j$ denote "$x_j \in U$ is true at time $t_j$," with $x_o = x$ and $x_n = x'$, repeated application of II.b gives

$$(x_o \to x_n) = \int_{x_1} \ldots \int_{x_{n-1}} (x_o \to x_1)(x_1 \to x_2) \times \ldots \times (x_{n-1} \to x_n). \tag{15}$$

As shown in reference 1, by repeating the same argument, each interval $(x_j \to x_{j+1})$ can be expanded into a sub-path integral which can then be reduced to a convolution by letting $t_j \to t_{j+1}$ and inverted with a fourier transform. As a result, for small $t' - t$, $|x' - x|$, the propagator $(x_t \to x'_{t'})$ is

$$(x_t \to (x+z)_{t+\tau}) = \frac{1}{(2\pi\tau)^{\frac{d}{2}}\sqrt{\det[\nu]}} \exp(-\tau \left[ \frac{1}{2}(\frac{z_j}{\tau} - \nu_j)W_{jk}(\frac{z_k}{\tau} - \nu_k) + \nu_0 \right]) \tag{16}$$

where $\nu_o(x)$, $\nu_j(x)$, $\nu_{jk}(x)$ and $W_{jk}(x)$ are moments of $\mu(x,z,\tau) \equiv (x_t \to (x+z)_{t+\tau})$ defined by $\nu_0(x) = \int_{z \in U} \mu_\epsilon(x,z,0)$, $\nu_j(x) = \int_{z \in U} \mu_\epsilon(x,z,0)z_j$, and $\nu_{jk}(x) = \int_{z \in U} \mu_\epsilon(x,z,0)z_j z_k$, with $W_{jk} = M_{jl}M_{lk}^T \omega_l$ where $M$ is the matrix which diagonalizes $\nu_{jk}$ such that $M_{lj}^T \nu_{jk} M_{km} = \delta_{lm}/\omega_l$. With velocity $v_j$ given by the limit of $z_j/\tau$, the above propagator is equivalent to the Lagrangian

$$\mathcal{L}(x,v) = \frac{i}{2}(v_j - \nu_j)W_{jk}(v_k - \nu_k) - i\nu_0 \tag{17}$$

which, by renaming the various moments, produces the Schrödinger equation for $d = 3$ or the Klein–Gordon equation for $d = 4$ (with $t$ identified as the proper time) where the particle mass, static vector and scalar potentials and metric ($W_{jk}$) all appear as moments of $(x_t \to x'_{t'})$. Notice that in contrast to the usual procedure, we have not assumed that the action is given by a classical Lagrangian or that the theory is Lorenz or gauge invariant.

Quantum theories also provide a convenient way to incorporate assumptions about an experiment directly into complex probabilities even if a solution to the full dynamics is not available. Consider, for example, the simple interferometer depicted in figure 1 where a photon encounters a beam splitter ($S_1$) followed by a mirror ($M_1$ or $M_2$) and a second beam splitter ($S_2$) thus reaching detector $D_1$ or $D_2$ via path $P_1$ and $Q_1$ or via path $P_2$ and $Q_2$. To simplify matters, ignore the photon polarization and consider a quantum theory with $U = R^3$. Let $e$ represent the initial description of the apparatus and photon, let $P_j$ = "The photon is on the path $P_j$ at a time $t$ after the first beam splitter is encountered but before the mirror is encountered," and similarly for $Q_j$ and let proposition $D_j$ be true if the photon reaches some chosen point at the entrance to detector $D_j$. Suppress time subscripts for convenience. Then, assuming that $D_j$ implies both $P_1 \vee P_2$ and $Q_1 \vee Q_2$, $(e \to D_j) = (e \to (P_1 \vee P_2) \wedge (Q_1 \vee Q_2) \wedge D_j)$ and using $P_1 \wedge P_2 = Q_1 \wedge Q_2 = false$,

$$(e \to D_j) = \sum_{n,m} (e \to P_n)(e \wedge P_n \to Q_m)(e \wedge P_n \wedge Q_m \to D_j). \tag{18}$$

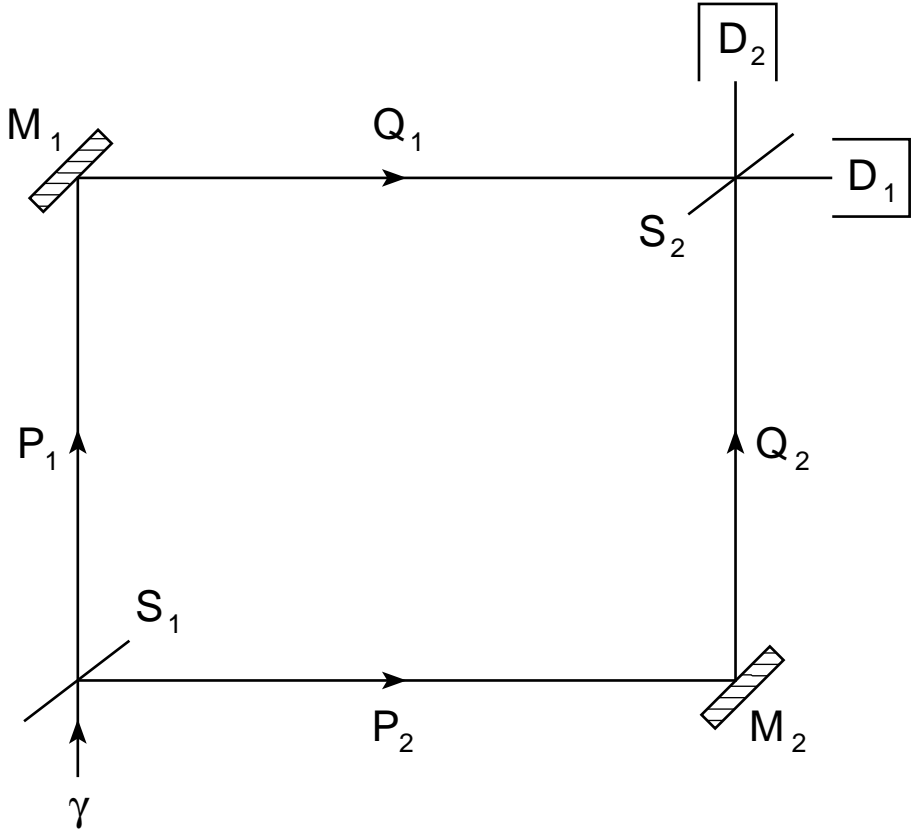

*Figure 1. A photon interferometer as described in the text where a single photon ($\gamma$) encounters a beam splitter ($S_1$), one of two mirrors ($M_1$ or $M_2$) and a second beam splitter ($S_2$) before entering one of two detectors ($D_1$ or $D_2$). The photon may either reach its detector by path $P_1$ and $Q_1$ or by path $P_2$ and $Q_2$.*

If $P_n$ and $Q_m$ are single points in $U$, then using II.c and assuming that $(P_n \to Q_m)$ is zero unless $n = m$, $(e \to D_j)$ is given by the sum of the complex probabilities for the two paths

$$(e \to D_j) = \sum_{n=1}^{2} (e \to P_n)(P_n \to Q_n)(Q_n \to D_j). \tag{19}$$

Assume that $(e \to P_n) = k_1 e^{i\lceil n=2 \rceil \pi/2}$, $(P_n \to Q_n) = e^{i\pi/2}$ and $(Q_n \to D_j) = k_2 e^{i\lceil n \neq j \rceil \pi/2}$ where $k_1$ and $k_2$ are complex constants and where $\lceil a \rceil$ denotes the function which is 1 if $a$ is *true* and 0 if $a$ is *false*. Thus, $(e \to D_1) = 0$ by cancellation of the two paths and since $Prob(e, D_1)$ is proportional to $|(e \to D_1)|^2$, there will be no counts observed in detector $D_1$. Suppose, now, that the mirror $M_1$ is attached to a spring and to a device $H$ which has two states: "hit" if the oscillation in the spring has ever been above some threshold and "nohit" otherwise. Equation (18) follows as before, but (19) does not since $P_n$ is no longer a point in the enlarged state space $U' = R^3 \times \{\text{hit}, \text{nohit}\}$. Use subscripts $e$, $P$, $Q$ and $D$ to indicate times, so that $\text{hit}_Q$ denotes "H is in state 'hit' at the time when Q is evaluated" etc., and let $e' = e \wedge \text{nohit}_e$. By the definition of $Prob$ in the new state space $U'$, $Prob(e', D_1)$ is proportional to $|e' \to D_1 \wedge \text{hit}_D|^2 + |e' \to D_1 \wedge \text{nohit}_D|^2$. If $H$ is completely successful at measuring whether a photon has struck $M_1$, then we can assume that $\text{hit}_D \Rightarrow Q_1$ and $\text{nohit}_D \Rightarrow Q_2$ and $\text{hit}_D \Rightarrow \text{hit}_Q \Rightarrow \text{nohit}_P$. Thus,

$$(e' \to D_1 \wedge \text{hit}_D) = (e' \to P_1 \wedge \text{nohit}_P \wedge Q_1 \wedge \text{hit}_Q \wedge D_1 \wedge \text{hit}_D)$$
$$= (e' \to P_1 \wedge \text{nohit}_P)(P_1 \wedge \text{nohit}_P \to Q_1 \wedge \text{hit}_Q)(Q_1 \wedge \text{hit}_Q \to D_1 \wedge \text{hit}_D) \quad (20)$$

and since there are no longer two terms to interfere, $Prob(e', D_1)$ is no longer zero. At the other extreme, if $H$ is far from sensitive enough to determine if $M_1$ has been struck by the photon, we can assume that $\text{hit}_t$ and $\text{nohit}_t$ are independent of $P_j$, $Q_j$ and $D_j$ for all $t$. Then $(e' \to D_1 \wedge \text{hit}_D)$ is equal to

$$\sum_{h,h' \in \{\text{hit,nohit}\}} (e' \to h_P)(h_P \to h'_Q)(h'_Q \to \text{hit}_D) \sum_{n=1}^{2} (e' \to P_n)(P_n \to Q_n)(Q_n \to D_1) \quad (21)$$

and thus $(e' \to D_1 \wedge \text{hit}_D) = (e' \to D_1 \wedge \text{nohit}_D) = 0$ and the previous prediction of no hits in $D_1$ is restored, as expected. The same mechanism for removing irrelevant propositions by assuming independence is also the source of the "locality" of quantum theories. For example, if two unrelated experiments $e_1$ and $e_2$ have possible results $r_1$ and $r_2$ respectively, then the assumptions that $\{r_1, r_2\}$, $\{e_1, r_2\}$ and $\{e_2, r_1\}$ are independent allow the expected conclusion $(e_1 \wedge e_2 \to r_1 \wedge r_2) = (e_1 \to r_1)(e_2 \to r_2)$.

## 5. Basic Properties of Quantum Theories

In section 4, we have seen that an energy eigenfunction should not be viewed as a state of the system, but rather as the best description of a system given certain initial knowledge. In particular, if only $U_o$ is known, an energy eigenstate is the best description of the system in a conservative, time invariant quantum theory. In effect, energy eigenstates arise from a nearly complete lack of knowledge of where the system is in the state space. As might be expected, eigenfunctions of other operators arise in a similar way. Given a quantum theory $U$ with normal initial knowledge $e_o$ and given a Hermitian operator $Q$, let

$$q(V) = \frac{\int_V \Psi^*(x) Q \Psi(x)}{\int_V \Psi^*(x) \Psi(x)} \quad (22)$$

where $V \subset U$ and $\Psi(x) = (e_o \to x_o)$. If $\Psi$ and $Q\Psi$ are continuous and if there is no reason to choose $q(V) \neq q(V')$ for any $V, V' \subset U$, $\Psi$ must be chosen from among the eigenfunctions of $Q$. If $Q\phi_j = \lambda_j \phi_j$ defines the eigenfunctions of $Q$, let

$$b_j = \text{"The system, at time } t = 0 \text{, is best described by } \phi_j\text{"} \quad (23)$$

so that $b_j \wedge b_k = false$ if $j \neq k$ and, by assumption, $\vee_j b_j = true$. Therefore

$$(e_o \to x_o) = (e_o \to \vee_j b_j \wedge x_o) = \sum_j (e_o \to b_j)(e_o \wedge b_j \to x_o) = \sum_j (e_o \to b_j)\phi_j(x) \quad (24)$$

which provides an expansion of the initial wavefunction in eigenfunctions of Q. Since $Q$ is Hermitian, its eigenfunctions are orthogonal, and, by theorem 1, $\{b_1, b_2, \ldots\}$ supports probabilities with initial knowledge $e_o$ and so

$$Prob(e_o, b_j) = |e_o \to b_j|^2 \frac{\int_U |\phi_j(x)|^2}{\int_U |e_o \to x_o|^2} \quad (25)$$

has a frequency interpretation. To put this expansion in a more familiar form, let $a_j = (e_o \to b_j)(\int_U |\phi_j(x)|^2 / \int_U |e_o \to x_o|^2)^{\frac{1}{2}}$, $\Psi(x) = (e_o \to x_o)/(\int_U |e_o \to x_o|^2)^{\frac{1}{2}}$ and $\psi_j(x) = \phi_j(x)/(\int_U |\phi_j(x)|^2)^{\frac{1}{2}}$ so that $\Psi(x) = \sum_j a_j \psi_j(x)$ where the $\psi_j(x)$ are eigenfunctions of Q and where $Prob(e, b_j) = |a_j|^2$. We have thus derived what is usually taken as the expansion postulate in conventional quantum mechanics.

In conventional quantum mechanics, a sharp distinction is made between pure states, which can be described by a single wavefunction and statistical mixtures, which must, in general, be described by a density matrix. Since probability theory itself is no longer available to us, these "statistical mixtures" must be described entirely within complex probability theory. To investigate this issue, consider several situations which require density matrices in conventional theory. First, consider a system with initial knowledge $e_o$ which is known to be well described by one of the wavefunctions $\psi_1, \psi_2, \ldots$ which may or may not be orthogonal. This would normally be represented as a mixture. As before, we have $(e_o \to x_o) = \sum_j (e_o \to b_j)\psi_j(x)$ where $b_j$ = "The system at $t = 0$ is best described by $\psi_j$." Thus, in a quantum theory, not knowing which $\psi_j$ best describes a system is no different from a pure superposition of $\psi_j$. To put it another way, all such expansions can be considered as mixtures with, in general, complex probabilities as coefficients and where a "statistical" mixture is only a special case. Density matrices are also needed in the case of "open systems" where $S \subset U$ is "the system" and $R = U - S$ is the rest of the world. If $e_S$ summarizes the initial knowledge of the system $S$, then with $x \in S$, $\psi(x) = (e_S \to x_o)$ is the initial wavefunction, as usual. In quantum theory, the single function $\psi(x)$ is sufficient to describe $S$ independent of what is known about the rest of the world if, for propositions $e_R$ about the rest of the world, $(e_S \wedge e_R \to x_o) = (e_S \to x_o)$ which is just the condition that $e_R$ and $x_o$ are independent. As a final example, consider a composite quantum theory $U = U_a \times U_b$ where Hermitian operator $\tilde{A}$ satisfies $\tilde{A}\phi_j = \tilde{a}_j \phi_j$ ($\phi_j : U_a \to C$) and Hermitian operator $\tilde{B}$ satisfies $\tilde{B}\psi_k = \tilde{b}_k \psi_k$ ($\psi_k : U_b \to C$). Following reference 10, evaluate the expectation value of $\tilde{A}$. For convenience, let all propositions implicitly be evaluated at time $t = 0$ and define $a_j$ = "$U_a$ is best described by $\phi_j$" and $b_k$ = "$U_b$ is best described by $\psi_k$." With initial normal knowledge $e$, $\{a_1, a_2, \ldots\}$ supports probabilities by virtue of theorem 1 and so we can define the expectation value $<\tilde{A}> = \sum_j \tilde{a}_j Prob(e, a_j)$. Using the definition of $Prob$,

$$<\tilde{A}> = \sum_j \tilde{a}_j \int_{x_a \in U_a} \int_{x_b \in U_b} (e \to a_j \wedge x_a \wedge x_b)^* (e \to a_j \wedge x_a \wedge x_b)/Z \qquad (26)$$

where $Z = \int_{U_a} \int_{U_b} |e \to x_a \wedge x_b|^2$. Since $\vee_k b_k = true$ and using the orthogonality of $(e \to a_j \wedge b_k \wedge x_a \wedge x_b)$ in both indices,

$$<\tilde{A}> = \sum_{j,k,m} \int_{U_a} \int_{U_b} (e \to a_j \wedge b_k \wedge x_a \wedge x_b)^* \tilde{A}(e \to a_m \wedge b_k \wedge x_a \wedge y_b)/Z \qquad (27)$$

and so

$$<\tilde{A}> = \sum_k |e \to b_k|^2 \int_{U_a} \int_{U_b} (e \wedge b_k \to x_a \wedge x_b)^* \tilde{A}(e \wedge b_k \to x_a \wedge x_b)/Z. \qquad (28)$$

Let the wavefunction "relative to $b_k$" be defined by $\Psi^k_{rel}(x_a, x_b) = (e \wedge b_k \to x_a \wedge x_b)/(\int_{U_a} \int_{U_b} |e \wedge b_k \to x_a \wedge x_b|^2)^{\frac{1}{2}}$. Then

$$<\tilde{A}> = \sum_k Prob(e, b_k) \int_{x \in U} \Psi^{k*}_{rel}(x) \tilde{A} \Psi^k_{rel}(x) \qquad (29)$$

agrees with the result from reference 10 where a demonstration is also given that no single wavefunction defined on $U_a$ gives the correct marginal distributions for all operators defined on $U_a$.

From these examples, we conclude that quantum theories are able to naturally describe mixtures without requiring extension of the axioms. Of course, in the case of the spin of an electron, a density matrix rather than a superposition is required to represent, for example, an unpolarized particle. Although we do not treat spin here, this observation indicates that spin cannot be described by a quantum theory with a two element state space.

There has been special recent interest in precision tests of quantum mechanics and in the question of whether non–linear extensions of quantum mechanics are possible.[11] Since linearity is assumed in conventional quantum mechanics, it is not possible to investigate this question within the usual framework. We can, however, show that linearity is a consequence of axioms I-III. Let proposition $e$ be the description of an experiment and let $E$ be a set of propositions describing a set of possible initial conditions. For convenience, suppress a $t = 0$ subscript on all propositions and let $\Psi_a(x) = (e \wedge a \to x)$ for any proposition $a$. Introduce $\tilde{a} = $ "The system is best described by $\Psi_a$" and $\tilde{b} = $ "The system is best described by $\Psi_b$" and let $z(e, E, a, b) = \{((e \wedge e' \to \tilde{a}), (e \wedge e' \to \tilde{b})) : e' \in E\}$. Linearity of quantum theories is then guaranteed by the following.

**Theorem 3 (Linearity).** *If $e$ is a proposition and $E$ is a set of propositions and $U$ is a quantum theory, then, for any pair $(\alpha, \beta) \in z(e, E, a, b)$ such that $\alpha + \beta \neq 0$, there exists a proposition $c$ such that $(\alpha + \beta)\Psi_c(x) = \alpha \Psi_a(x) + \beta \Psi_b(x)$.*

*Proof.* Let $(\alpha, \beta) \in z(e, E, a, b)$ and let $e'$ satisfy $(e \wedge e' \to \tilde{a}) = \alpha$ and $(e \wedge e' \to \tilde{b}) = \beta$. The case where $\tilde{a} = \tilde{b}$ is trivial. Otherwise, let $w = e \wedge e' \wedge (\tilde{a} \vee \tilde{b})$. We have $(w \to x) = (w \to (\tilde{a} \vee \tilde{b}) \wedge x)$ [since $(e \wedge e' \to \tilde{a} \vee \tilde{b}) = \alpha + \beta \neq 0$] which equals $(w \to \tilde{a} \wedge x) + (w \to \tilde{b} \wedge x)$ which equals $(w \to \tilde{a})\Psi_a(x) + (w \to \tilde{b})\Psi_b(x)$. Since $(w \to \tilde{a}) = \alpha/(\alpha + \beta)$ and $(w \to \tilde{b}) = \beta/(\alpha + \beta)$, the theorem is proven if we let $c = e' \wedge (\tilde{a} \vee \tilde{b})$.

Since the proof of theorem 3 only uses the complex probability axioms, the linearity of quantum theories is a direct result of the addition of probabilities in I.b and so a non–linear extension of quantum mechanics would not be consistent with complex probability theory or with Cox's axioms.

## 6. The Bayesian View of Complex Probabilities

One of the benefits of realistic quantum theories based on complex probability is a simple explanation of the puzzling, not quite paradoxical problems in quantum mechanics involving the collapse of the wavefunction, non–local effects and the role of the observer in the theory. Within realistic quantum theories, all of these problems are ultimately due to mistaking the wavefunction for the state of the system as discussed in section 4. Here we illustrate how the idea that a wavefunction represents knowledge about a system and not it's state[12] arises in the Bayesian view of complex probabilities.

The puzzling phenomenon of wavefunction collapse can be illustrated by imagining a wave packet which strikes a barrier causing part of the wave packet to be reflected to the left and part to be transmitted to the right. After some time, a box on the right is sealed and a device tests (with 100% efficiency) whether there is a particle in the box. If the test is successful, a repeated measurement must also be successful with probability 1, and so the wave packet on the left must, instantaneously and mysteriously, travel to the box on the right, penetrate its walls and join the right half of the wave packet. However, from the point of view of a realistic quantum theory, after striking the barrier, the particle did either go to the left or the right and the diverging wave packets only represent the fact that we do not know which way it actually went. The collapse of the wavefunction just corresponds to learning a new fact about the system and using that fact in calculating new complex probabilities. Since this is only a change in the description of the system, it does not correspond to anything actually happening to the particle, the box or to the observer. Similarly, in the EPR experiment as described above in the discussion of Bell's theorem, the measurement of one of the spins has no effect on the remote particle and so, for instance, it should not be possible to use EPR correlations for non–local communication.[13] This difference between viewing the wavefunction as a representation of what one knows about a system rather than its state is directly analogous to the difference between the classical view of a probability as something determined by an underlying random phenomenon and the Bayesian view where a probability represents what one happens to know about a system.[6]

The same point can be illustrated once again by Schrödinger's cat, who is put in an opaque box with a device which will kill it with probability $\frac{1}{2}$. In the mysterious version, after the device has acted, it would not be correct to say that the cat is either alive or dead until an observer opens the box and looks, causing the wavefunction of the cat to collapse into one of the two alternatives. However, consider, for comparison, the classical problem of a six sided die in an opaque box. If the upward face of the die is unknown, a Bayesian would describe this situation with a "superposition" $p_j = 1/6$, $j = 1, \ldots, 6$ (determined, for instance, by maximizing entropy) and there would thus be a "collapse of the probability distribution" when the box is opened and the die face is revealed. However, the "superposition" $p_j$ and collapse is not mysterious at all precisely because $p_j$ is not mistaken for the "state of the die." Similarly, the mystery of Schrödinger's cat disappears if one takes the Bayesian view of complex probabilities.

Since, as we have seen, wavefunctions only represent what is known about a system rather than its state, quantum theory should also be useful for systematically improving wavefunctions based on prior information in the same sense that Bayesian inference[6] is used to improve probability distributions. Analogues of both components of Bayesian inference are available in quantum theories since Bayes theorem follows [if $(a \to b) \neq 0$, then $(a \wedge b \to c) = (a \to c)(a \wedge c \to b)/(a \to b)$] and since a suitable maximum entropy principle has already been proposed and successfully used in many–body problems.[14]

## 7. Summary

Starting with complex probability theory and a simple statement of realism, we are able to derive much of what is assumed in conventional quantum mechanics including the probability interpretation, the superposition principle, the expansion postulate and the wave equation including static fields and a metric. These quantum theories have a convenient treatment of mixed systems which are represented without extension to the axioms. The Bayesian view of complex probabilities provides an easy understanding of the EPR experiment, Bell's theorem and the other mysterious problems in conventional quantum mechanics and suggests a program for improving wave functions analogous to Bayesian inference in probability theory. It remains to be seen whether particles with spin, multi–particle systems, bosons and fermions and field theories can also be based on complex probability theory. Open questions about the fundamentals of quantum theories also remain. In particular, the sense in which probability theory is restored in the classical limit needs to be quantified. It might also be possible to show that the state axioms are a consequence of the complex probability axioms in the sense that they are required for a frequency interpretation to exist.

R.N. Silver in *Workshop on Physics and Computation (PhysComp92)* (IEEE Computer Society Press, 1993).